\documentstyle[twocolumn,prl,aps,psfig]{revtex} 
\newcommand{ \be }{\begin{equation}}
\newcommand{ \ee }{\end{equation}}
\newcommand{ \bea }{\begin{eqnarray}}
\newcommand{ \eea }{\end{eqnarray}}
\newcommand{ \la }{\langle}
\newcommand{ \ra }{\rangle}
\newcommand{ \eps }{{\varepsilon}}

\begin{document}

\title{
\vspace*{-2.2cm}
\begin{flushright}
{{\small \sl version 20 \,
\today}}
\end{flushright}
Elliptic flow in Au+Au collisions at $\sqrt{s_{_{NN}}}=130$~GeV}
\author{
\small
K.H.~Ackermann$^{19}$, N.~Adams$^{28}$, C.~Adler$^{12}$,
Z.~Ahammed$^{27}$,  S.~Ahmad$^{28}$, C.~Allgower$^{13}$, J.~Amsbaugh$^{34}$, 
M.~Anderson$^6$, E.~Anderssen$^{17}$, H.~Arnesen$^3$
% ...
\\
(STAR Collaboration)
}

\address{
% ...
$^3$Brookhaven National Laboratory,
% ...
$^6$University of California, Davis,
% ...
$^{12}$University of Frankfurt,
$^{13}$Indiana University,
% ...
$^{17}$Lawrence Berkeley National Laboratory,
% ...
$^{19}$Max-Planck-Institut fuer Physik, Munich,
% ...
$^{27}$Purdue University,
$^{28}$Rice University,
% ...
$^{34}$University of Washington,
% ...
}

\date{\today}

\maketitle

\begin{abstract}
Elliptic flow from nuclear collisions is a hadronic observable
sensitive to the early stages of system evolution. We report first
results on elliptic flow of charged particles at midrapidity in Au+Au
collisions at $\sqrt{s_{_{NN}}}=130$~GeV using the STAR TPC at
RHIC. The elliptic flow signal, $v_2$, averaged over transverse
momentum, reaches values of about $6\%$ for relatively peripheral
collisions and decreases for the more central collisions. This can be
interpreted as the observation of a higher degree of thermalization
than at lower collision energies. Pseudorapidity and transverse
momentum dependence of elliptic flow are also presented.
\end{abstract}

\vspace*{-1.0cm}

The goal of the ultrarelativistic nuclear collision program is the
creation of a system of deconfined quarks and gluons~\cite{qm99}.  If
this system is created, its evolution should be governed by the
physics of deconfined matter.  The elliptic flow observable, which is
sensitive to the early evolution of the system, is the anisotropic
emission of particles ``in'' or ``out'' of the reaction plane defined
for non-central collisions by the beam direction (z-axis) and the
impact parameter direction (x-axis). Elliptic flow is usually
characterized in terms of particle momenta by $v_2 = \la (p_x^2 -
p_y^2)/(p_x^2 + p_y^2) \ra$, the second harmonic Fourier coefficient
in the azimuthal distribution of particles with respect to the
reaction plane~\cite{VZ,meth}.  Elliptic flow has its origin in the
spatial anisotropy of the system when it is created in a non-central
collision, and in particle rescatterings in the evolving system which
convert the spatial anisotropy to momentum anisotropy.  The spatial
anisotropy in general decreases with system expansion, thus quenching
this effect and making elliptic flow particularly sensitive to the
early stages of the system evolution~\cite{sorge98}.  Being dependent
on rescattering, elliptic flow is sensitive to the degree of
thermalization of the system~\cite{vpPLB,heinz00} at this early
time. Hydrodynamic models, which are based on the assumption of
complete local thermalization, predict the strongest
signals~\cite{heinz00,olli92,sollfrank,shuryak}.

Elliptic flow in ultrarelativistic nuclear collisions was first
discussed in Ref.~\cite{olli92} and has been studied intensively in
recent years at AGS~\cite{e877flow2,e895} and
SPS~\cite{na49prl,na49flow,wa98} energies. These studies have found
that elliptic flow at high energies is ``in-plane'', $v_2>0$, as
expected from most models, and the pion elliptic flow for relatively
peripheral collisions increases with beam energy~\cite{olli98} from
about 2\% at the top AGS energy~\cite{e877flow2} to about 3.5\% at the
SPS~\cite{na49flow}. From transport cascade models for the full RHIC
energy, a peak elliptic flow value of 1.5\% is predicted by
UrQMD~\cite{UrQMD} calculations and 2.5\% by RQMD v2.4~\cite{sorge95}
calculations~\cite{STARnote}. Hydrodynamic models predict $v_2$ as
high as 10\%~\cite{heinz00,sollfrank}. Details of the $v_2$ dependence
on beam energy and centrality are thought to be sensitive to the phase
transition between confined and deconfined
matter~\cite{vpPLB,heinz00,sollfrank,shuryak,heisel99} (see
also~\cite{olli98} and references therein).

We report here the first results on elliptic flow in Au+Au collisions
at the Relativistic Heavy Ion Collider (RHIC) at
$\sqrt{s_{_{NN}}}=130$~GeV.  The Solenoidal Tracker At RHIC
(STAR)~\cite{STAR} consists of several detector sub-systems in a large
solenoidal magnet.  For first year data taking, the setup consists of
the Time Projection Chamber (TPC) which covers the pseudorapidity
range $|\eta| < 1.8$ for collisions in the center of the TPC, and has
complete azimuthal coverage, which is desirable for the study of
azimuthal correlations.  In the first year, the TPC is operated with a
0.25 Tesla field, allowing tracking of particles with
$p_t>75$~MeV/c. Two Zero Degree Calorimeters~\cite{ZDC} which measure
fragmentation neutrons are used in coincidence for the trigger.  The
TPC is surrounded by a scintillator barrel which measures the charged
particle multiplicity, and is used in studies of the trigger
performance and vertex reconstruction efficiency.

\begin{figure}
\centerline{\psfig{figure=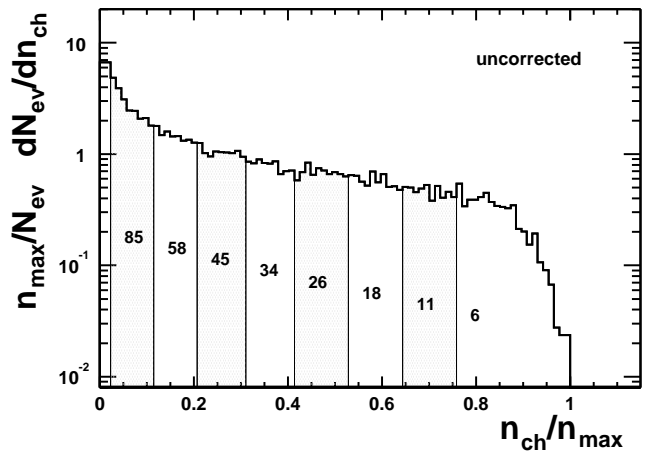,height=6.7cm}}
  \caption[]{ The primary track multiplicity distribution as a
  function of the number of tracks normalized by the maximum observed
  number of tracks. The eight centrality regions used in this analysis
  are shown. The integral under the curve is 1.0 and the cumulative
  fraction corresponding to the lower edge of each centrality bin is
  also indicated in percent. }
\label{mult}
\end{figure}

The relative multiplicity distribution for events with a reconstructed
primary vertex is shown in Fig.~\ref{mult}. An analysis of the trigger
performance and vertex reconstruction efficiency, together with
comparisons with Hijing~\cite{hijing}, show that the events in
Fig.~\ref{mult} are hadronic Au+Au interactions corresponding to about
90\% of the geometric cross section, the losses being due to vertex
reconstruction inefficiency for low-multiplicity events. This vertex
finding inefficiency is not included in the normalization of
Fig.~\ref{mult}. The multiplicity is the number of primary tracks
which pass within 3 cm of the vertex and have $|\eta| < 0.75$.  The
distribution shown is not corrected for tracking efficiency; it
is used in this analysis only to estimate centralities.

For this analysis, 22~k events were selected with a primary vertex
position within 75~cm longitudinally of the TPC center and within 1~cm
radially of the beam line. Tracks were selected with $0.1 < p_t \le
2.0$ GeV/c in order to have a tracking efficiency constant to within
$\pm~10\%$. They also passed within 1~cm of the primary vertex, had at
least 15 space points, and $|\eta| < 1.3$. For the determination of
the event plane we required $|\eta| < 1.0$. Also, the ratio of the
number of space points to the expected maximum number of space points
for that particular track was required to be greater than 0.52,
largely suppressing split tracks from being counted twice. However,
the analysis results are not sensitive to these cuts.

\begin{figure}
\centerline{\psfig{figure=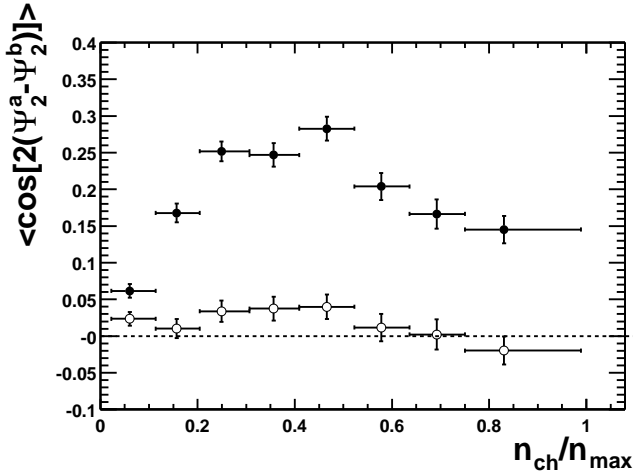,height=6.5cm}}
  \caption[]{ Correlation between the event plane angles determined
  for two independent subevents. The upper set of data is for the
  second harmonic and the lower set is $\la \cos(\Psi^a-\Psi^b)\ra$
  for the first harmonic. The points are positioned at the values of
  mean $n_{ch}/n_{max}$ corresponding to each of the centrality bins
  in Fig.\ref{mult}. The horizontal bars show the widths of the
  bins. }
\label{corr}
\end{figure}

The analysis method~\cite{VZ,meth} involves the calculation of the
event plane angle, which is an experimental estimator of the real
reaction plane angle. The second harmonic event plane angles,
$\Psi_2$, are calculated for two subevents, which are groups of
independent particles from the same event. In order to see whether
these planes are correlated, the mean cosine of the difference in
their event plane angles is calculated.  Although the STAR
detector has good azimuthal symmetry, small acceptance effects in the
calculation of the event plane angle were removed by the methods of
shifting or weighting~\cite{meth}. This correction, by either method,
is negligible for the second harmonic. 

The subevents have been chosen in three different ways: 1) Assigning
particles with pseudorapidity $0.05<\eta<1$ to one subevent and
particles with $-1<\eta<-0.05$ to the other subevent.  The ``gap''
between the two regions ensures that short range correlations, such as
Bose-Einstein correlations or Coulomb final state interactions,
contribute negligibly to the observed correlation. 2) Dividing all
particles randomly into two subevents. 3) Assigning positive particles
to one subevent and negative particles to the other.  Fig.~\ref{corr}
shows the results for correlation of the event planes of subevents
assigned by the pseudorapidity method. The two other methods give
similar results.  Non-flow effects (not correlated with the reaction
plane) would contribute differently for these different subevent
choices. The shape of the centrality dependence of the second-harmonic
signal is characteristic of anisotropic flow and quite different from
possible non-flow sources.

Most commonly discussed non-flow sources of azimuthal correlations
are: 1) Momentum conservation, which can affect directed flow when
each subevent is not symmetric about mid-rapidity, does not affect
elliptic flow measurements.  2) Coulomb and Bose-Einstein
correlations~\cite{olli00a}, which are eliminated by the construction
of the subevents in Fig~\ref{corr}.  3) Resonance
decay~\cite{olli00b}, whose effect on the subevent correlation would
be independent of centrality, unlike what is observed for the second
harmonic.  4) Jets, when calculated using Hijing~\cite{hijing} for the
cuts used in the current analysis, do not contribute beyond the
systematic errors for $v_2$ quoted below. Also, if jets or resonances
contribute to $v_2$ they would contribute to the directed flow
measurements in comparable amount\cite{foothijing}. The first harmonic
correlation, which is shown in Fig~\ref{corr}, and the higher
harmonics, are about an order of magnitude weaker than the elliptic
flow at mid-centrality. This sets an upper limit for the contribution
of all non-flow effects to the elliptic flow and is the basis of the
estimate below of the systematic errors.

The analysis method involves correlation of the azimuthal angle,
$\phi$, of each particle with an event plane angle, $\Psi$, and then
averaging over all events. In this paper we have used three particle
correlation methods: 1) Correlating the particles from one hemisphere
with the event plane of the subevent in the other hemisphere. 2)
Correlating each particle with the event plane of all the {\sl other}
particles. 3) Correlating particles of one charge sign with the event
plane of the opposite charge sign.  The observed elliptic flow comes
from the second harmonic Fourier coefficient of the particle azimuthal
distribution with respect to the event plane, which is simply $\la
\cos[2(\phi-\Psi_2)]\ra$. The elliptic flow relative to the real
reaction plane, $\Psi_R$, the plane defined by the impact parameter
and the beam, can be evaluated by dividing the observed signal by the
resolution, $\la \cos[2(\Psi -
\Psi_R)]\ra$, of the event plane. The resolutions calculated from the
correlation of subevent planes were somewhat different for the
different subevent selections, but the resultant $v_2$ values were the
same within statistical errors.  The resolutions for the full events
reach 0.7 for the centrality at the peak in Fig.~\ref{corr}, while in
NA49\cite{na49prl,na49flow} at the SPS they only reached 0.4. A
resolution of the event plane angle of 0.7 is sufficiently close to
the ideal value of 1.0, to correlate other quantities, such as
two-particle correlation measurements (HBT) with the event
plane. Since we do not measure the correlation with the first harmonic
plane, we cannot determine the sign of $v_2$.

Our analysis procedures have been tested with simulated
data\cite{mevsim} to which a known amount of flow has been added. The
simulated data were filtered by a GEANT model of STAR and
reconstructed in a way similar to that used for the data.  For 2\% and
10\% elliptic flow added to the simulations, the flow extracted was
$(2.0 \pm 0.1)\%$ and $(9.7 \pm 0.2)\%$, respectively.

\begin{figure}
\centerline{\psfig{figure=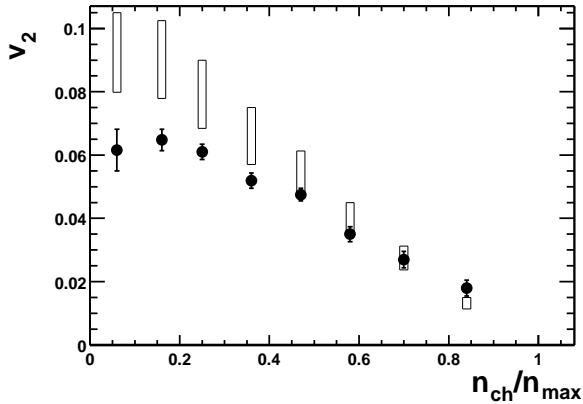,height=6.0cm}}
  \caption[]{ Elliptic flow (solid points) as a function of centrality
  defined as $n_{ch}/n_{max}$. The open rectangles show a range of
  values expected for $v_2$ in the hydrodynamic limit, scaled from
  $\epsilon$, the initial space eccentricity of the overlap
  region. The lower edges correspond to $\epsilon$ multiplied by 0.19
  and the upper edges to $\epsilon$ multiplied by 0.25. }
\label{cen}
\end{figure}

Fig.~\ref{cen} shows $v_2$ as a function of centrality of the
collision.  Although this figure was made with the subevents chosen as
in Fig.~\ref{corr}, the same results within errors were obtained with
the other correlation methods. Restricting the primary vertex $z$
position to reduce TPC acceptance edge effects also made no
difference. From the results of the study of non-flow contributions by
different subevent selections and the maximum magnitudes of the first
and higher-order harmonics, we estimate a systematic error for $v_2$
of about 0.005, with somewhat smaller uncertainty for the
mid-centralities where the resolution of the event plane is high. The
systematic errors are not included in the figures.

In the hydrodynamic limit, elliptic flow is approximately proportional
to the initial space anisotropy, $\epsilon$, which is calculated in
Ref.~\cite{jacobs99}. The transformation to the multiplicity axis in
Fig.~\ref{cen} was done using a Hijing~\cite{hijing} simulation,
taking into account the above mentioned vertex-finding inefficiency for
low multiplicity events. In comparing the flow results to $\epsilon$,
no unusual structure is evident which could be attributed to the
crossing of a phase transition while varying
centrality\cite{sorge98,heisel99}. The $\epsilon$ values in
Fig.~\ref{cen} are scaled to show the range of hydrodynamic
predictions\cite{heinz00,sollfrank} for $v_2/\epsilon$ from 0.19 to
0.25. The data values for the lower multiplicities could indicate
incomplete thermalization during the early time when elliptic flow is
generated~\cite{vpPLB,heinz00}. On the other hand, for the most
central collisions, comparison of the data with hydrodynamic
calculations suggest that early-time thermalization may be
complete. The $v_2$ values peak at more peripheral collisions than
RQMD predictions~\cite{STARnote}, but in qualitative agreement with
hydrodynamic models~\cite{olli92}.

\begin{figure}
\centerline{\psfig{figure=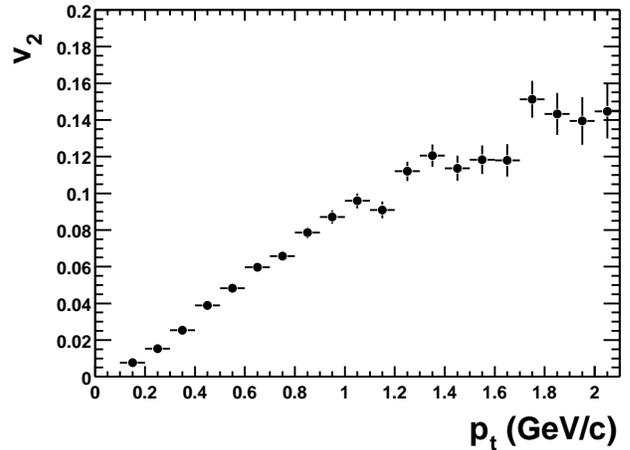,height=6.5cm}}
  \caption[]{ Elliptic flow as a function of transverse momentum for
  minimum bias events.  }
\label{pt}
\end{figure}

The differential anisotropic flow is a function of $\eta$ and
$p_t$. For the integrated results presented here, all $v$ values
should first be calculated as a function of $\eta$ and $p_t$, and then
averaged over either or both variables using the double differential
cross sections as weights. Since we do not yet know the cross
sections, we have averaged using the observed yields. Fig.~\ref{pt}
shows $v_2$ as a function of $p_t$ for a minimum bias trigger. The
$\eta$ dependence (not shown), which is averaged over $p_t$ from 0.1
to 2.0 GeV/c, is constant at a value of $(4.5 \pm 0.5)$\% for $|\eta|
\lesssim 1.3$.  We have assumed that the efficiency (yield/cross
section) is constant in the $p_t$ range where the yield is large.
This is borne out by studies of the effects of different track quality
cuts on the observed $p_t$ spectra.  For the $p_t$ dependence the data
are not very sensitive to the assumption of constant efficiency as a
function of $\eta$ because $v_2$ appears to be independent of $\eta$
in the range used, $|\eta| < 1.3$.  Mathematically the $v_2$ value at
$p_t=0$, as well as its first derivative, must be zero, but it is
interesting that $v_2$ appears to rise almost linearly with $p_t$
starting from relatively low values of $p_t$.  This is consistent with
a stronger ``in-plane'' hydrodynamic expansion of the system than the
average radial expansion. Note that the results shown in
Fig.~\ref{cen} were obtained by taking the average over both $\eta$
and $p_t$, weighted by the yield. Although Fig.~\ref{pt} is for
approximately minimum bias data\cite{footmb} the general shapes are
the same for data selected on centrality, except that the slopes of
the $p_t$ curves depend on centrality. Fig.~\ref{pt} was made using
pseudorapidity subevents, although the same results within errors were
obtained using the other two methods.

We conclude that elliptic flow at RHIC rises up to about 6\% for the
most peripheral collisions, a value which is more than 50\% larger
than at the SPS~\cite{na49flow}, indicating stronger early-time
thermalization at this RHIC energy. Elliptic flow
appears to be independent of pseudorapidity in the region $|\eta|
\lesssim 1.3$. Its $p_t$ dependence is almost linear in the region
$0.1 < p_t < 2$~GeV/c.  Comparing to estimates~\cite{STARnote} based
on transport cascade models, we find that elliptic flow is
underpredicted by RQMD by a factor of more than 2. Hydrodynamic
calculations~\cite{heinz00,sollfrank} for RHIC energies overpredict
elliptic flow by about 20-50\%. This is just the reverse of the
situation at the SPS where RQMD gave a reasonable description of the
data and hydrodynamic calculations were more than a factor of two too
high~\cite{na49flow}. Also in contrast to lower collision energies,
the observed shape of the centrality dependence of the elliptic flow
is similar to hydrodynamic calculations and thus consistent with
significant thermalization. The values for elliptic flow compared to
hydrodynamic models indicate that early-time thermalization is
somewhat incomplete for peripheral collisions but perhaps complete for
the more central collisions.

{\bf Acknowledgments:} We wish to thank the RHIC Operations Group at
Brookhaven National Laboratory for their tremendous support and for
providing collisions for the experiment. This work was supported by
the Division of Nuclear Physics and the Division of High Energy
Physics of the Office of Science of the U.S.Department of Energy, the
United States National Science Foundation, the Bundesministerium
f\"{u}r Bildung und Forschung of Germany, the Institut National de la
Physique Nucleaire et de la Physique des Particules of France, the
United Kingdom Engineering and Physical Sciences Research Council, and
the Russian Ministry of Science and Technology.

\noindent
\\$^{*}$Deceased


\begin{thebibliography}{99}
\bibitem{foothijing}
  Jet {\sl quenching} could depend on the orientation of the jet with
  respect to the reaction plane.  This would be a real flow
  correlation.  Hijing predicts~\cite{STARnote} the contribution of
  jet quenching to $v_2$ to be small at mid-rapidity for particles
  with $p_t < 2$~GeV/c.
\bibitem{footmb}
  We say approximately because the quality cuts discriminate somewhat
  against the more central collisions. At each of the eight
  centralities, $v_2$ was calculated and the values were combined by
  weighting with the inverse of the square of the errors, which is
  proportional to the number of particles in that bin. This should be
  the same as a theoretical calculation of $\la \cos(2\phi) \ra$ for
  all particles in all events.
\end{thebibliography}
\end{document}